\documentclass[twocolumn,aps,prl,superscriptaddress]{revtex4-1}
\usepackage{graphicx}
\usepackage{amsmath}
\usepackage{bm}
\usepackage[squaren,Gray]{SIunits}

\begin{document}

\title{Giant Collective Spin-Orbit Field in a Quantum Well: Fine Structure\\ of Spin Plasmons}

\author{F. Baboux}
\thanks{Corresponding author: \texttt{florent.baboux@insp.upmc.fr}}
\affiliation{Institut des Nanosciences de Paris, CNRS/Universit\'e Paris VI, Paris 75005, France}

\author{F. Perez}
\affiliation{Institut des Nanosciences de Paris, CNRS/Universit\'e Paris VI, Paris 75005, France}

\author{C. A. Ullrich}
\affiliation{Department of Physics and Astronomy, University of Missouri, Columbia, Missouri 65211, USA}

\author{I. D'Amico}
\affiliation{Department of Physics, University of York, York YO10 5DD, United Kingdom}

\author{J. G\'omez}
\altaffiliation{Present address: Centro At\'omico Bariloche, Bariloche, Argentina.}
\affiliation{Institut des Nanosciences de Paris, CNRS/Universit\'e Paris VI, Paris 75005, France}

\author{M. Bernard}
\affiliation{Institut des Nanosciences de Paris, CNRS/Universit\'e Paris VI, Paris 75005, France}

\begin{abstract}
We employ inelastic light scattering with magnetic fields to study intersubband spin plasmons in a quantum well.
We demonstrate the existence of a giant collective spin-orbit (SO) field that splits the spin-plasmon spectrum into a
triplet. The effect is remarkable as each individual electron would be expected to precess in its own momentum-dependent
SO field, leading to D'yakonov-Perel' dephasing. Instead, many-body effects
lead to a striking organization of the SO fields at the collective level. The macroscopic spin moment is quantized by a uniform collective SO field, five times higher than the individual SO field.
We provide a momentum-space cartography of this field.
\end{abstract}

\maketitle

%%%%%%%%%%%%%%%%%%%%%%%%%%%%%%%%%%%%%%%%%%%%%%%%%%%%%%%%%%%%%%%%%%%%%%%%%%%%%%%%%%%%%%%%%%%%%%%%%%%%%%%%%%%%%%%%%%%%%%%%%%%%%%%%%%%%%%%%%%%%%%%%%%%%%%%%%%%%%%%%%%%%%%%%%%%%%%%%%%%%%%%%%%%%%%%%%%%%%%%%%%%%%%%%%%%%%%%%%%%%%%%%%%%%%%%%%%%%%%%%%%%

Spin-orbit (SO) coupling arises from relativity: the spin of
an electron moving at a velocity $\mathbf{v}$ in a static electric field $%
\mathbf{E}$ sees a magnetic field $\mathbf{B}_{\mathrm{SO}}=-\frac{1}{c^{2}}%
\mathbf{v}\times \mathbf{E}$ ($c$ is the speed of light)~\cite{Thomas1926}.
This magnetic field splits the energy levels of atoms, giving rise to their fine structure~\cite{Lamb1950}.
For an ensemble of itinerant electrons in solids, such a simple quantizing effect cannot be expected
because of the distribution of velocities. Momentum-dependent SO fields cause each individual electronic spin
to precess with its own axis, which destroys spin coherence
(D'yakonov-Perel' [DP] decoherence~\cite{Dyakonov1971}). This
sets practical limitations on many proposed
applications in emerging quantum technologies such as spintronics~\cite{Zutic2004,Sih2005,Kato2004b,Chernyshov2009,Koralek2009}.

However, this DP picture is appropriate only for situations where the macroscopic spin is carried by
individual electrons, which is often the case~\cite{Das1989,Jusserand1992,Sih2005,Meier2007,Koralek2009}.
Here, we demonstrate that Coulomb interaction, which plays a central role in \textit{collective}
spin excitations, can drastically modify this picture, and give rise to macroscopic quantum objects.
We will focus on intersubband spin plasmons in doped
semiconductor quantum wells, which, as we shall see, are ideal to study the interplay of SO coupling and Coulomb interactions.

In a III-V quantum well, internal SO fields arise from the lack of an inversion center of the crystalline unit cell,
and from an asymmetric confining potential~\cite{WinklerBook}, referred to as Dresselhaus~\cite{Dresselhaus1955} and
Rashba~\cite{Rashba1984} fields, respectively.
Hence, a conduction electron with momentum $\mathbf{k}$,
moving in the plane of a $\left[ 001\right] $-oriented quantum well, experiences a SO magnetic field
\begin{equation}
\mathbf{B}_{\mathrm{SO}}(\mathbf{k})=\frac{2\alpha }{g\mu _{\mathrm{B}}}%
\begin{pmatrix}
k_{y} \\
-k_{x}%
\end{pmatrix}%
+\frac{2\beta }{g\mu _{\mathrm{B}}}%
\begin{pmatrix}
k_{x} \\
-k_{y}%
\end{pmatrix}
\label{BSO}
\end{equation}%
(to lowest order in $\mathbf{k}$), for coordinate systems with $\hat x \parallel \nolinebreak \left[ 100\right]$
and $\hat y \parallel \left[ 010\right]$. Here, $\alpha $ and $\beta $ are the Rashba and linear Dresselhaus
coupling constants~\cite{WinklerBook}, respectively, $g$ is the electron g-factor, and $\mu _{\mathrm{B}}$ the Bohr magneton. $\mathbf{B}_{\mathrm{SO}}$ produces
an intrinsic $\mathbf{k}$-dependent spin splitting~\cite{Das1989,Jusserand1992}
and a $\mathbf{k}$-dependent spin orientation~\cite{Dyakonov1971b,Kato2004b} of single-electron conduction states.

In such a system, electrons can exhibit collective spin dynamics when excited from
the first to the second subband of the quantum well.
These so-called intersubband (ISB) spin plasmons, which arise from Coulomb interactions, are energetically well
separated from the continuum of ISB single-particle excitations~\cite{Pinczuk1989,Gammon1990}.
In the absence of a transferred momentum
$\mathbf{q}$ and external magnetic field $\mathbf{B}_{\mathrm{ext}}$, time
reversal symmetry, together with the $\left[ 001\right] $-axis symmetry of
the quantum well, average out the $\mathbf{k}$-dependent $\mathbf{B}_{\mathrm{SO}}$.
Hence, no macroscopic SO force is acting on the electron gas, and the spin plasmons are degenerate. However, when
transferring an in-plane momentum $\mathbf{q}$ to the electron gas, the translation symmetry is broken and $\mathbf{B}_{%
\mathrm{SO}}(\mathbf{k})$ does not average out anymore.

\begin{figure}[tbp]
\includegraphics[width=\columnwidth]{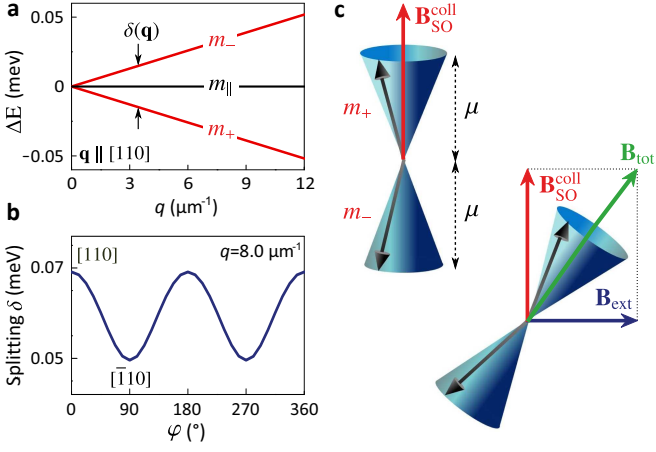}
\caption{Fine structure model of ISB spin plasmons. (a) Threefold splitting,
induced by SO coupling, of the ISB spin plasmon modes. $\Delta E$
denotes the difference of the mode energies with and without SO coupling,
calculated for the studied GaAs quantum well, and $q$ is the magnitude of the plasmon momentum
(here $\mathbf{q}\parallel \left[ 110\right] $). The splitting $\delta $ between the transverse $m_{\pm }$ modes is
almost linear in $q$. (b) For a fixed $q=\unit{8.0}{\reciprocal{\micro%
\metre}}$, calculated modulation of the splitting $\delta$ with the
in-plane orientation of $\mathbf{q}$, labeled by the angle $\varphi$
to $\left[ 110\right]$. (c) Sketch of the proposed interpretation of
the transverse ISB spin plasmons $m_{\pm }$, as the precession of
antiparallel $\mu $ collective magnetic moments about $\mathbf{B}_{%
\mathrm{SO}}^{\mathrm{coll}}$ at zero external field (left), and about
the superposition $\mathbf{B}_{\mathrm{SO}}^{\mathrm{coll}}\mathbf{+B_{%
\mathrm{ext}}}$ when an external magnetic field $\mathbf{B_{\mathrm{ext}}}$
is applied (right).}
\label{Theory}
\end{figure}

In this situation, it has been predicted \cite{Ullrich2002,Ullrich2003} that
despite the spread of $\mathbf{B}_{\mathrm{SO}}(\mathbf{k})$, a collective
SO magnetic field $\mathbf{B}_{\mathrm{SO}}^{\mathrm{coll}}(\mathbf{q%
})$ emerges, splitting the spin plasmon branch into three modes [Fig.~\ref{Theory}(a)]:
one longitudinal oscillation mode ($m_{\parallel }$) and two
transverse precession modes ($m_{+}$ and $m_{-}$).
In the present work, we focus on these
transverse modes, whose frequencies are shifted in opposite directions by
SO coupling. We propose that due to Coulomb
interaction, these modes behave as \textit{macroscopic quantum objects},
characterized by a collective spin magnetic moment $\mathbf{M}$ and, thus, subject to an interaction energy $W\mathbf{(q)}=-\mathbf{M}\cdot \mathbf{B}%
_{\mathrm{SO}}^{\mathrm{coll}}(\mathbf{q})$. Within this framework, the
downward (upward) energy shift of the $m_{+}$ ($m_{-}$) mode is explained
by its projected magnetic moment being parallel (antiparallel) to the
quantizing field $\mathbf{B}_{\mathrm{SO}}^{\mathrm{coll}}(\mathbf{q})$
[Fig.~\ref{Theory}(c), left]. Then, in the presence of an external magnetic
field, we expect both fields to superpose~\cite{Sih2005,Meier2007} [Fig.~\ref{Theory}(c), right] and the interaction energy to become
\begin{equation}
W\mathbf{(q)}=-\mathbf{M}\cdot \left( \mathbf{B}_{\mathrm{SO}}^{\mathrm{coll}%
}(\mathbf{q})\mathbf{+B_{\mathrm{ext}}}\right) .  \label{W}
\end{equation}%
We will demonstrate that this fine structure model correctly
captures the physics of the ISB spin plasmons.

\begin{figure}[tbp]
\includegraphics[width=\columnwidth]{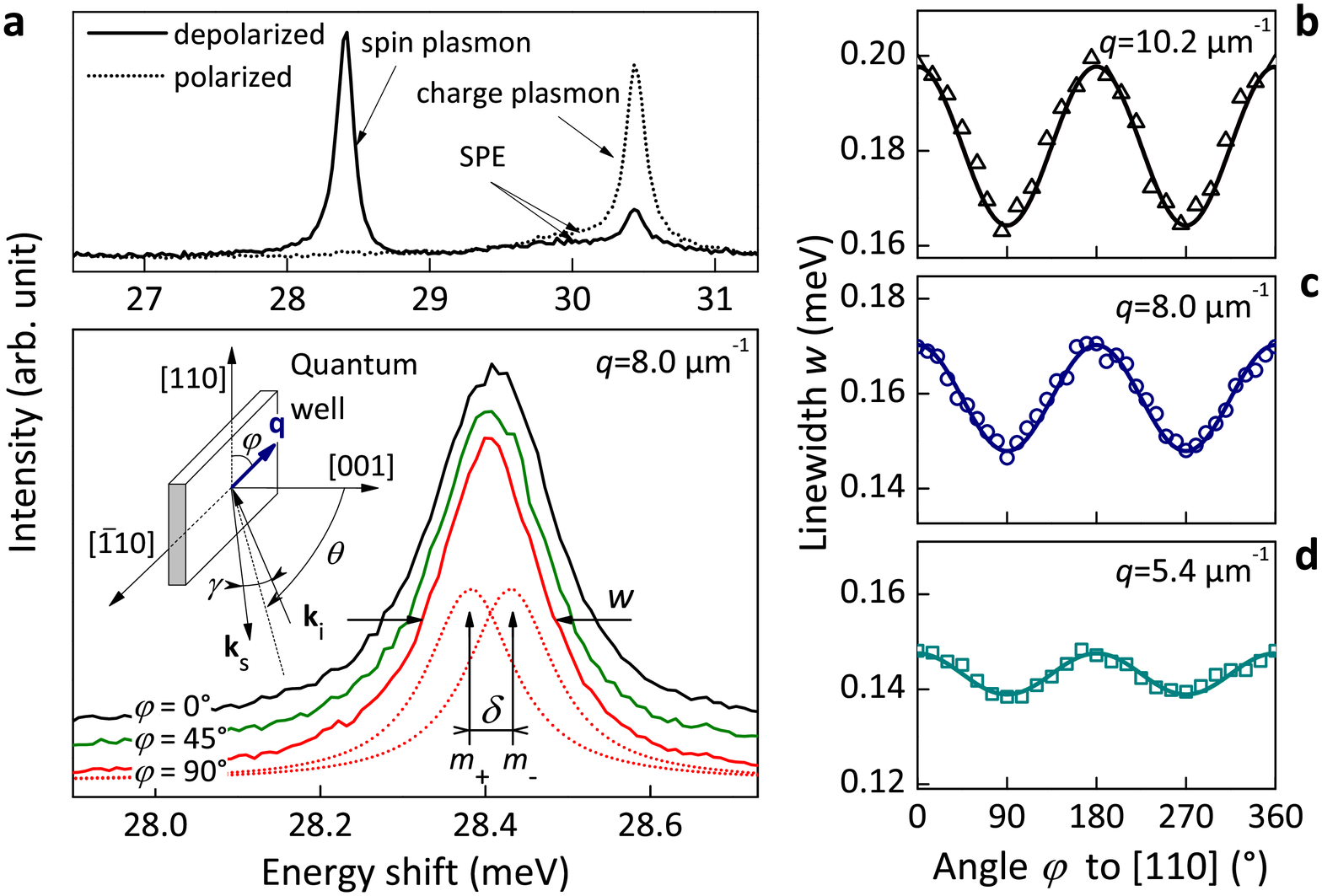}
\caption{Anisotropic splitting of the ISB spin plasmon
modes. (a)
Top panel: Inelastic light scattering spectrum of the ISB excitations, in polarized (dashed line) and depolarized (solid line) geometry.
Bottom panel:
Depolarized spectra obtained at fixed $q=\unit{8.0}{\
\reciprocal{\micro\metre}}$, by varying the in-plane angle $\varphi $
measured from $\left[ 110\right] $ (vertical offset for clarity). The
single, quasi-Lorentzian peak observed is the sum of two Lorentzians (red
dashed lines for the $\varphi =\unit{90}\!{\degree}$ spectrum) of
same amplitude and linewidth, corresponding to the
transverse spin plasmons modes $m_{+}$ and $m_{-}$ split by an
amount $\delta$. Inset: scattering geometry showing angle definitions; $\mathbf{k_{i}}$ and $\mathbf{%
k_{s}}$ are the incoming and scattered light wavevectors.
(b-d) Variation of the linewidth $w$
with $\varphi $ for $q=10.2,\;8.0\;\text{and}\;\unit{5.4}{\
\reciprocal{\micro\metre}}$ respectively.
Lines: Theory (see text).}
\label{Spectra}
\end{figure}

We carry out inelastic light scattering (ILS) measurements in a $[001]$ oriented, asymmetrically modulation-doped
GaAs/AlGaAs quantum well. The electron density is $\unit{2.3\times 10^{11}}{\rpsquare{\centi\metre}}$, and the mobility $\unit{2\times 10^{7}}{%
\squaren{\centi\metre}\,\reciprocal{\volt}\,\reciprocal{\second}}$ at the
working temperature $T\simeq \unit{2}{\kelvin}$ (superfluid helium)~\cite{SuppM}.
ILS~\cite{Pinczuk1989,Perez2007} is a powerful tool to study spin excitations at a given transferred
momentum $\mathbf{q}$ [Fig.~\ref{Spectra}(a), inset]~\cite{SuppM}.
Standard selection rules~\cite{Pinczuk1989} allow us to address the various
types of intersubband excitations individually. As shown in the
spectra of Fig.~\ref{Spectra}(a) (top), the charge plasmon is observed only when the
incident and scattered photon have parallel polarizations (polarized geometry), while the spin plasmon appears
when they have orthogonal polarizations (depolarized geometry). The single-particle
excitations continuum appears in both configurations (here as a shoulder of the charge plasmon peak).
From now on we focus on the spin plasmon peak, obtained in the depolarized geometry where only the transverse modes, $m_{+}$
and $m_{-}$ are probed.
Typical spectra, taken in the absence of an external
magnetic field, are presented in Fig.~\ref{Spectra}(a) (bottom).
These are obtained for
a momentum of fixed magnitude $q=\unit{8.0}{\reciprocal{\micro\metre}}$, but
various in-plane orientations, labeled by the angle $\varphi $ between $%
\mathbf{q}$ and the $\left[ 110\right] $ direction of the
quantum well.
The spectra exhibit a single, quasi-Lorentzian peak of full width
at half-maximum (FWHM) $w$; when plotting $w$ for various $\varphi$
[Fig.~\ref{Spectra}(c)], $w$ is modulated quasi-sinusoidally with a period $%
\pi $. This modulation is characteristic of the twofold symmetry
of the SO splitting [Fig.~\ref{Theory}(b)], with a maximum along $%
\left[ 110\right] $ ($\varphi =\unit{0}\!{\degree}$) and a minimum along $%
\left[ \overline{1}10\right] $ ($\varphi =\unit{90}\!{\degree}$). Furthermore,
as seen in Figs.~\ref{Spectra}(b)-(d), the amplitude of the modulation decreases
with decreasing $q$, in agreement with Fig.~\ref{Theory}(a). Both
characteristics confirm the SO origin of the modulation. This suggests
that the observed Raman line is the sum of two Lorentzian peaks corresponding
to the transverse spin plasmon modes $m_{+}$ and $m_{-}$, split by $\delta $ [Fig.~\ref{Spectra}(a), red dashed
lines]. By independently determining the FWHM of the latter peaks, we will extract
the splitting $\delta (\mathbf{q})$ by deconvolution, and demonstrate the consistency of our model.

We determine the collective SO field $\mathbf{B}_{\mathrm{SO}}^{\mathrm{%
coll}}(\mathbf{q})$ by applying an external magnetic
field $\mathbf{B_{\mathrm{ext}}}$. Since $\mathbf{B}_{\mathrm{SO}}^{\mathrm{%
coll}}(\mathbf{q})$ is expected~\cite{Ullrich2003} to lie in the plane of the quantum well for
our $\left[ 001\right] $-oriented sample, $\mathbf{B_{\mathrm{ext}}}$ will be applied in the well plane (quasi-Voigt geometry).
If the spin plasmon moment $\mathbf{M}$ is of quantum nature, its energy levels will be
quantized by the total field $\mathbf{B}_{\mathrm{tot}}=\mathbf{B}_{\mathrm{%
SO}}^{\mathrm{coll}}(\mathbf{q})\mathbf{+B_{\mathrm{ext}}}$. Following Eq.~(\ref{W}), the
splitting $\delta$ will then be given by
\begin{equation}
\delta =2 \mu B_{\mathrm{tot}}=2 \mu \sqrt{(B_{\mathrm{ext}}+\mathbf{B}_{%
\mathrm{SO}}^{\mathrm{coll}}\cdot \mathbf{u}){}^{2}+(\mathbf{B}_{\mathrm{SO}%
}^{\mathrm{coll}}\times \mathbf{u}){}^{2}},  \label{delta}
\end{equation}%
where $\mu$ is the quantized value of the spin plasmon magnetic moment and $\mathbf{u}$ is a unit vector parallel to the direction of $\mathbf{B_{\mathrm{ext}}}$.

\begin{figure*}[tbp]
\centering
\includegraphics[width=\textwidth]{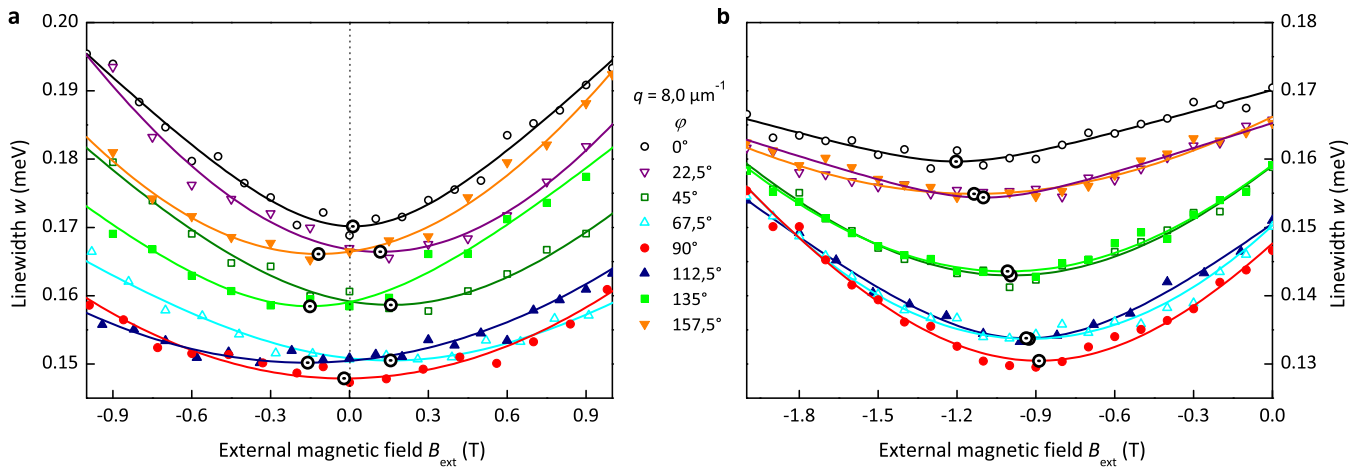}
\caption{Variation of the composite linewidth $w$ with the external
magnetic field $\mathbf{B}_{\mathrm{ext}}$. (a) $w(B_{\mathrm{ext}})$ plots obtained in the
configuration $\mathbf{B_{\mathrm{ext}}\parallel q}$ for a fixed $q=\unit{8.0%
}{\ \reciprocal{\micro\metre}}$ and various in-plane angles $\varphi
$ (measured from $\left[ 110\right] $). (b) Corresponding $w(B_{%
\mathrm{ext}})$ plots obtained for $\mathbf{B_{\mathrm{ext}}\perp
q}$. Lines are guides for the eyes. Each $w(B_{\mathrm{ext}})$ plot is
symmetric about a certain value of the external
field (marked by a dotted circle) which cancels the corresponding component of the
collective SO field $\mathbf{B}_{\mathrm{SO}}^{\mathrm{coll}}(%
\mathbf{q})$. }
\label{Bext}
\end{figure*}

For a given $\mathbf{q}$, we record a series of spectra at varying $B_{\mathrm{ext}}$, with
$\mathbf{B_{\mathrm{ext}}}$ applied successively along two crossed
directions: $\mathbf{B_{\mathrm{ext}}}\perp \mathbf{q}$ and $\mathbf{B_{%
\mathrm{ext}}}\parallel \mathbf{q}$. Figures~\ref{Bext}(a) and \ref{Bext}(b),
respectively, present the \textit{composite} linewidth $w$ as a function of $B_{\mathrm{ext}}$
for these two configurations. The various plots are obtained
for fixed $q=\unit{8.0}{\reciprocal{\micro\metre}}$, and a set of eight
angles $\varphi $, spaced by $\unit{22.5}\!{\degree}$ within a period $\pi $.
Each plot exhibits a clear minimum for a certain value of $B_{\mathrm{ext}}$,
and is symmetric with respect to that minimum.

\begin{figure}
\centering
\includegraphics[width=\columnwidth]{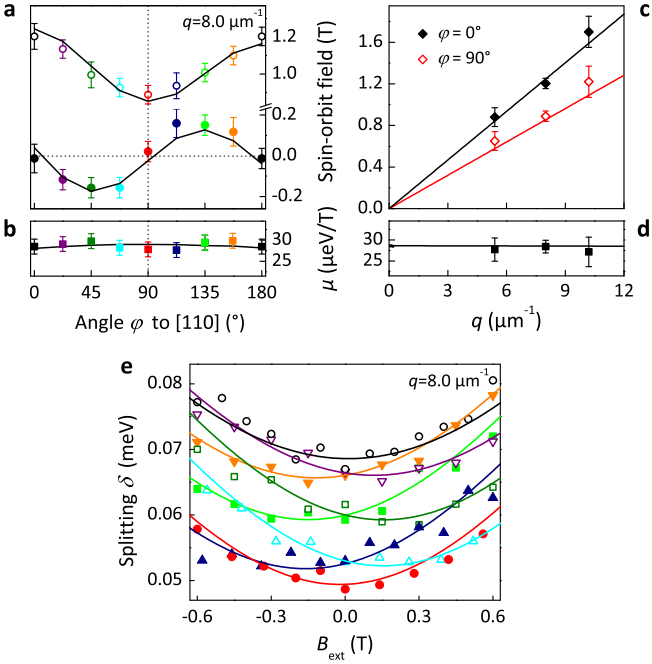}
\caption{SO collective field and magnetic moment. (a) Components of the collective SO field parallel ($B_{\mathrm{SO}%
,\parallel }^{\mathrm{coll}}$, filled circles) and perpendicular ($B_{%
\mathrm{SO},\perp }^{\mathrm{coll}}$, open circles) to $\mathbf{q}$ for $q=%
\unit{8.0}{\ \reciprocal{\micro\metre}}$, as extracted from the data of Fig.~%
\ref{Bext}, and compared with theory (lines). (b) Spin plasmon magnetic moment $\mu $, experimental (squares)
and theoretical (line).
(c) Minimum (open diamonds) and maximum (filled
diamonds) SO field $\left\vert \mathbf{B}_{\mathrm{SO}}^{\mathrm{coll%
}}\right\vert $ versus $q$,
compared to theoretical values (lines). (d) Spin plasmon
magnetic moment averaged over $\varphi $, experimental (squares) and theoretical (line),
as a function of $q$. (e) Variation of the SO splitting $\delta $ with external
magnetic field $\mathbf{B_{\mathrm{ext}}\parallel q}$, for $q=\unit{8.0}{\
\reciprocal{\micro\metre}}$ (symbols, same as Fig.~\ref{Bext}).
The experimental data are very well reproduced by Eq.~(\ref{delta}) (lines).
}
\label{Field}
\end{figure}

According to Eq.~(\ref{delta}), each minimum corresponds to the situation where $\mathbf{B}_{\mathrm{ext}}$ exactly cancels
the component of $\mathbf{B}_{\mathrm{SO}}^{\mathrm{coll}}(\mathbf{q})$
parallel to it, $B_{\mathrm{ext}}=-\mathbf{B}_{\mathrm{SO}}^{\mathrm{coll}%
}\cdot \mathbf{u}$. Using this criterion, we extract the component $B_{%
\mathrm{SO,}\parallel }^{\mathrm{coll}}$ of the collective SO field
parallel to $\mathbf{q}$ from the plots of Fig.~\ref{Bext}(a) and the
perpendicular component $B_{\mathrm{SO,}\perp }^{\mathrm{coll}}$ from the
plots of Fig.~\ref{Bext}(b). Figure~\ref{Field}(a) presents the values for $B_{%
\mathrm{SO,}\parallel }^{\mathrm{coll}}$ (filled circles) and $B_{\mathrm{SO,%
}\perp }^{\mathrm{coll}}$ (open circles).
We find that $B_{\mathrm{SO,}\parallel }^{\mathrm{coll}}$ is antisymmetric about the $%
\left[ \overline{1}10\right] $ direction ($\varphi =\unit{90}\!{\degree)}$
and $B_{\mathrm{SO,}\perp }^{\mathrm{coll}}$ is symmetric.

To push the analysis further, we need experimental access to the SO splitting $\delta $. This can be done by determining the
linewidth of the $m_{\pm }$
modes [see Fig. \ref{Spectra}(a)]. The latter is inferred
from the zero external field and zero momentum value of the
FWHM $w$ of the composite peak (not shown), since in that case we expect the splitting $\delta$ to vanish [see Fig.~%
\ref{Theory}(a)] and both peaks to lie perfectly on top of each other. This yields $0.124\pm \unit{0.005}{\milli\electronvolt}$.

This linewidth value can be compared to theory. ISB spin plasmons
are expected to be immune against DP dissipation~\cite{Ullrich2002,Ullrich2003}. Thus, owing
to the very high mobility of the sample
and the low working temperature,
we expect the linewidth to be dominated by an intrinsic many-body effect, the spin
Coulomb drag~\cite{Damico2000,Damico2006,Weber2005} (SCD).
The SCD is caused by a friction of Coulomb origin between carriers of opposite spin moving with different momenta.
The ISB spin plasmon, where spin densities oscillate out of phase along the growth axis, provides an optimal scenario
for the SCD~\cite{Damico2006}. A calculation of the corresponding
linewidth within a local-density approximation yields an SCD linewidth of the
order of a fraction of an ${\milli \electronvolt}$, confirming that the dominant dissipation
source is the SCD. As the latter is mainly due to the out-of-plane spin density oscillation,
its $\mathbf{q}$ dependence, for $\mathbf{q}$ much smaller than the Fermi momentum, is weak
and only to second order. Hence, we deconvolute all the $w(B_{\mathrm{ext}})$ curves using the
experimentally determined $ \unit{0.124}{\milli\electronvolt}$.

Figure \ref{Field}(e) presents $\delta (B_{\mathrm{ext}})$ (symbols as in Fig.~\ref{Bext}),
obtained by deconvolution of the data of Fig.~\ref{Bext}(a). Using Eq.~(\ref{delta}),
we can now evaluate the collective magnetic moment of the spin plasmons
as $\mu=\delta \left( B_{ \mathrm{ext}}=0\right) / \left(2 \left\vert \mathbf{B}_{\mathrm{SO}}^{\mathrm{coll}}\right\vert\right)$.
This ratio is plotted in Fig.~\ref{Field}(b) (squares), for the various $\varphi$ probed. It appears constant with $\varphi $ within the experimental error. We deduce $\mu =28.8\pm \unit{0.7}{\micro\electronvolt\,\reciprocal{\tesla}}=\left( 0.50\pm 0.01\right) \mu _{\mathrm{B}}$.

The consistency of our interpretation of the data with the
model of Eq.~(\ref{delta}) is demonstrated in Fig.~\ref{Field}(e), which compares the experimental
data points for $\delta \left( B_{\mathrm{ext}}\right)$ with the relation
$\delta \left( B\mathbf{_{\mathrm{ext}}}\right) =2 \mu \sqrt{(B_{\mathrm{ext}}
+B_{\mathrm{SO},\parallel }^{\mathrm{coll}}){}^{2}+B_{\mathrm{SO},\perp }^{\mathrm{coll}}{}^{2}}$ (lines), using the previously determined values of
$B_{\mathrm{SO},\parallel }^{\mathrm{coll}}$, $B_{\mathrm{SO},\perp }^{\mathrm{coll}}$ and $\mu $. An excellent agreement is found, without introducing any fitting parameters.

We further validate our model by checking the $q$ dependence of $\mu $. We repeat the
same experimental procedure for other values of $q$. Figure~\ref{Field}(c) presents the values of the minimum
($\varphi =\unit{90}\!{\degree}$, open diamonds) and maximum
($\varphi =\unit{0}\!{\degree}$, filled diamonds) modulus of $\mathbf{B}_{\mathrm{SO}}^{\mathrm{%
coll}}$. They appear proportional to $q$. Figure~\ref{Field}(d) shows the
angular average of the magnetic moment $\mu $ (squares). Interestingly,
$\mu $ turns out to be practically constant with $q$. This demonstrates that
all of the SO effects are contained in $\mathbf{B}_{\mathrm{SO}}^{\mathrm{coll}}\mathbf{(q)}$,
and that $\mu $ is indeed the largest quantized projection of the intrinsic ISB spin plasmon magnetic moment $\mathbf{M}$ onto the
field direction.
This point is confirmed by noting that $\mu \approx 2\frac{\left\vert
g\right\vert \mu _{\mathrm{B}}}{2}$, that is, $\mu$ is very close to
twice the magnetic moment of a single electron (when considering the
g-factor of bulk GaAs, $g=-0.445$).
This is consistent with the fact that
an ISB spin plasmon involves transitions between two spin $1/2$ states, i.e.
excitations of spin magnitude $1$. Our results thus show that the ISB plasmon maintains
the spin magnitude of a single elementary excitation, while the many-body effects
are absorbed in the collective magnetic field $\mathbf{B}_{\mathrm{SO}}^{%
\mathrm{coll}}\mathbf{(q)}$.
Hence, the quantized projection of the plasmon magnetic moment
can either be $\pm \mu$ [$m_{\pm}$ modes, see Fig.~\ref{Theory}(c)] or $0$ [$m_{\parallel }$ mode, whose energy is unaltered by SO coupling].

To summarize the experimental part, we validated our fine structure model by demonstrating
the internal consistency of measurements---with and
without magnetic field---with Eqs.~(\ref{W}) and (\ref{delta}). We emphasize again that this did not involve any adjustable parameters.

All elements are now in place to see how the collective SO
magnetic field $\mathbf{B}_{\mathrm{SO}}^{\mathrm{coll}}$ emerges from the $%
\mathbf{k}$-dependent single-particle magnetic fields $\mathbf{B}_{\mathrm{SO%
}}(\mathbf{k})$ given by Eq.~(\ref{BSO}).
The ISB spin plasmon is a superposition of single-particle transitions from
momentum $\mathbf{k}$ in the first subband to $\mathbf{k}+\mathbf{q}$ in the
empty second subband.
Thus, each electron-hole pair experiences a crystal magnetic field
difference $\Delta\mathbf{B_{\mathrm{SO}}(k,q)}=\mathbf{B_{\mathrm{SO,2}}(k+q)}-\mathbf{B_{\mathrm{SO,1}}(k)}$,
given by
\begin{eqnarray}
\lefteqn{\hspace{-5mm}
\frac{g\mu_{\mathrm{B}}}{2}\Delta\mathbf{B_{\mathrm{SO}}(k,q)}
=
\alpha _{\mathrm{2}}%
\begin{pmatrix}
q_{y} \\
-q_{x}%
\end{pmatrix}%
+\beta _{\mathrm{2}}%
\begin{pmatrix}
q_{x} \\
-q_{y}%
\end{pmatrix}}
\nonumber\\
&&{}
\hspace{+11mm}+\left( \alpha _{\mathrm{2}}-\alpha _{\mathrm{1}}\right)
\begin{pmatrix}
k_{y} \\
-k_{x}%
\end{pmatrix}%
+\left( \beta _{\mathrm{2}}-\beta _{\mathrm{1}}\right)
\begin{pmatrix}
k_{x} \\
-k_{y}%
\end{pmatrix}%
,  \label{SOchange}
\end{eqnarray}%
where the subscript $n=1,2$ refers to the subband index. With $\alpha _{\mathrm{1}}=\unit{3.5}{\milli
\electronvolt\,\angstrom}$, $\alpha _{\mathrm{2}}=\unit{2.8}{\milli\electronvolt\,
\angstrom}$, $\beta _{\mathrm{1}}=\unit{0.22}{\milli\electronvolt\,\angstrom}$, and
$\beta _{\mathrm{2}}=\unit{0.79}{\milli\electronvolt\,\angstrom}$, we are able to reproduce the experimental data in Figs.~\ref{Spectra}(b)--(d) and Figs.~\ref{Field}(a)--(e) in a quantitatively accurate way (see lines),
using a linear-response formalism based on time-dependent
density-functional theory.

$\Delta\mathbf{B_{\mathrm{SO}}(k,q)}$ contains a $\mathbf{k}$-independent part, which is thus the
same for all electron-hole pairs, and a $\mathbf{k}$-dependent part.
The latter could have a disorganizing effect, causing DP dephasing. This is indeed what occurs for
single-particle spin dynamics~\cite{Das1989,Jusserand1992,Sih2005,Meier2007,Koralek2009}. But here, the $\mathbf{k}$-dependence turns out
to be exactly canceled by an additional, dynamical Coulombic contribution~\cite{Ullrich2003},
explaining how a uniform $\mathbf{B}_{\mathrm{SO}}^{\mathrm{coll}}\mathbf{(q)}$ can emerge.

In a simple scenario, one could expect $\mathbf{B}_{\mathrm{SO}}^{\mathrm{%
coll}}\mathbf{(q)}$ to be aligned with the $\mathbf{k}$-independent part of
$\Delta\mathbf{B_{\mathrm{SO}}(k,q)}$, with a slightly enhanced magnitude. But what is
found is that $\mathbf{B}_{\mathrm{SO}}^{\mathrm{coll}}\mathbf{(q)}=%
\frac{2 \times 5.25}{ g {\mu }_{\mathrm{B}}}%
\left(\overline{\alpha }q_{y}+\overline{\beta }q_{x},-\overline{\alpha }q_{x}-\overline{\beta }q_{y}%
\right)$ (within $3\%$), with $\overline{\alpha }=\left( \alpha _{\mathrm{1}}+\alpha _{\mathrm{2%
}}\right) /2$ and $\overline{\beta }=\left( \beta _{\mathrm{1}}+\beta
_{\mathrm{2}}\right) /2$. That is, many-body effects tilt the $\mathbf{k}
$-independent part of $\Delta\mathbf{B_{\mathrm{SO}}(k,q)}$, align it with the average
single-particle SO field difference, and amplify it by a factor of about five.

Such a magnification effect due to dynamical many-body interactions is quite remarkable.
At first glance, one would expect Coulomb-induced enhancements to be roughly of order $r_{s}$ (Wigner-Seitz radius),
which is $\simeq 1.3$ for the studied sample. On the other hand, recent experiments~\cite{Liu2008,Nedniyom2009} suggest
that the interplay of Coulomb and SO interactions could manifest in
a mutual boost, leading to significant enhancement of electronic spin splittings especially in low-dimensional systems~\cite{Agarwal2011}.

In conclusion, we have shown that many-body effects can produce a significant
departure from the single-particle picture of spin-orbit effects in crystals.
In an intersubband spin plasmon, despite the spread of electronic velocities, a well-organized spin dynamics emerges at the collective level. The electrons coherently precess about
a giant spin-orbit field, which gives rise to a fine structure of the spin-plasmon spectrum.
This effect, which might also play a role in other helical liquids~\cite{Raghu2010,Agarwal2011}, reveals novel
opportunities for magnetization control with collective spin-orbit fields.

\begin{acknowledgments}
F.B. and F.P. thank S.~Majrab for technical support and B.~Jusserand for fruitful
discussion. F.P. acknowledges funding from C'NANO IDF 2009 (SPINWAVEDYN)
and ANR 2007 (GOSPININFO). F.B. is supported by a Fondation CFM-JP Aguilar grant. I.D'A. acknowledges
support from EPSRC Grant No. EP/F016719/1 and I.D'A. and F.P. acknowledge support from Royal Society Grant No. IJP 2008/R1 JP0870232.
C.A.U. is supported by DOE Grant No. DE-FG02-05ER46213.
\end{acknowledgments}

%%%%%%%%%%%%%%%%%%%%%%%%%%%%%%%%%%%%%%%%%%%%%%%%%%%%%%%%%%%%%%%%%%%%%%%%%%%%%%%%%%%%%%%%%%%%%%%%%%%%%%%%%%%%%%%%%%%%%%%%%%%%%%%%%%%%%%%%%%%%%%%%%%
%Bibliography
%%%%%%%%%%%%%%%%%%%%%%%%%%%%%%%%%%%%%%%%%%%%%%%%%%%%%%%%%%%%%%%%%%%%%%%%%%%%%%%%%%%%%%%%%%%%%%%%%%%%%%%%%%%%%%%%%%%%%%%%%%%%%%%%%%%%%%%%%%%%%%%%%%

%

\end{document}